\documentclass[aps,prb,reprint,preprintnumbers,superscriptaddress,twocolumn,nofootinbib]{revtex4}
\usepackage{graphicx}
\usepackage{bm,amsmath}
\usepackage{dcolumn}
\usepackage{amsfonts,amssymb}
\usepackage[colorlinks=true, pdfstartview=FitV, linkcolor=red, citecolor=blue, urlcolor=blue]{hyperref}
\usepackage{enumitem}

\def\Z{\mathbb{Z}}

\newcommand{\be}{\begin{equation}}
\newcommand{\ee}{\end{equation}}

\newcommand{\ket}[1]{\left|#1\right\rangle}
\newcommand{\bra}[1]{\left\langle#1\right|}
\newcommand{\brkt}[2]{\left\langle#1|#2\right\rangle}
\newcommand{\braket}[3]{\left\langle#1\left|#2\right|#3\right\rangle}

\newcommand{\diff}{\mathrm{d}}
\newcommand{\p}{\partial}
\newcommand{\ve}{\varepsilon}

\newcommand{\up}{\uparrow}
\newcommand{\down}{\downarrow}
\newcommand{\im}{\mathrm{i}}
\newcommand{\bea}{\begin{eqnarray}}      
\newcommand{\eea}{\end{eqnarray}}

\newcommand{\CxPxT}{$C$-$P$-$T$ }

\begin{document}
\title{C-P-T anomaly matching in bosonic quantum field theory and spin chains}

\author{Tin Sulejmanpasic} 
\affiliation{Philippe Meyer Institute, Physics Department, \'Ecole Normale Sup\'erieure, PSL Research University, 24 rue Lhomond, F-75231 Paris Cedex 05, France}

\author{Yuya Tanizaki} 
\email{yuya.tanizaki@riken.jp}
\affiliation{RIKEN BNL Research Center, Brookhaven National Laboratory, Upton, NY 11973, USA}

\date{\today}

\vskip 2 cm

\begin{abstract}
We consider the $O(3)$ nonlinear sigma model with the $\theta$-term and its linear counterpart in 1+1D. The model has discrete time-reflection and space-reflection symmetries at any $\theta$, and enjoys the periodicity in $\theta\rightarrow \theta+2\pi$. At $\theta=0,\pi$ it also has a charge-conjugation $C$-symmetry. Gauging the discrete space-time reflection symmetries is interpreted as putting the theory on the nonorientable $\mathbb RP^2$ manifold, after which the $2\pi$ periodicity of $\theta$ and the $C$ symmetry at $\theta=\pi$ are lost. We interpret this observation as a mixed 't Hooft anomaly among charge-conjugation $C$, parity $P$,  and time-reversal $T$ symmetries when $\theta=\pi$. Anomaly matching implies that in this case the ground state cannot be trivially gapped, as long as $C$, $P$ and $T$ are all good symmetries of the theory. We make several consistency checks with various semi-classical regimes, and with the exactly solvable XYZ model. We interpret this anomaly as an anomaly of the corresponding spin-half chains with translational symmetry, parity and time reversal (but not involving the $SO(3)$-spin symmetry), requiring that the ground state is never trivially gapped, even if $SO(3)$ spin symmetry is explicitly and completely broken. We also consider generalizations to $\mathbb{C}P^{N-1}$ models and show that the $C$-$P$-$T$ anomaly exists for even $N$.
\end{abstract}

\maketitle

\section{Introduction}

Spin chains are often effectively described by nonlinear or linear sigma models in 1+1D. In particular, Haldane showed \cite{Haldane:1982rj, Haldane:1983ru} that a quantum Heisenberg anti-ferromagnet in the limit of a large spin $S$, reduces to the nonlinear sigma model in 1+1D with the target space $S^2$ (the Bloch sphere). 
This model is referred to, conventionally, as the $O(3)$ nonlinear sigma model. 
Importantly Haldane showed that a natural topological $\theta$-term given by $\theta=2\pi S$ arises from the microscopic model, and argued that  the integer and half-integer spin fall separately into two universality classes since the partition function of the $O(3)$ model is invariant under the shifts of $\theta$ by $2\pi$. Indeed the spin-1/2 Heisenberg antiferromagnet is exactly solvable by the Bethe ansatz, and is gapless, while the AKLT model~\cite{Affleck:1987vf} as well as numerous simulations of the $\theta=0$ nonlinear sigma model show a gapped spectrum, often referred to as the Haldane gap.

The different behavior of the two regimes can be traced back to the Lieb-Schultz-Mattis (LSM) theorem \cite{Lieb:1961fr, PhysRevLett.84.1535, Hastings:2003zx}, which states that in the half-integer spin case, spin chains either have to be gapless or break translational symmetry~\cite{Affleck:1986pq}. No similar statement exists for integer spin systems. The underlying reason of the LSM theorem is that the $SO(3)$ spin symmetry acts projectively on the half-integer spins. 

It has recently come to light that the LSM theorem manifests itself as 't~Hooft anomaly in the effective theory (see e.g. Refs.~\cite{Cho:2017fgz,Metlitski:2017fmd,Jian:2017skd} which touch upon this connection). The 't Hooft anomaly can be summarized as an inability to gauge a particular symmetry group without introducing non-local terms in the action. Unlike quantum anomalies, 't Hooft anomalies do not invalidate the global symmetry, but instead give powerful constraints on the IR physics. The 't Hooft anomaly matching condition was originally proposed for continuous chiral symmetry of gauge theories with massless fermions~\cite{tHooft:1979rat, Frishman:1980dq, Coleman:1982yg}, where it was used to predict that chiral symmetry must be broken or the system must be in the conformal phase. 

How 't Hooft anomalies constrain the infrared physics can be intuitively explained as follows (We review the anomaly matching more precisely in the Appendix \ref{app:anomaly_matching}). For the sake of simplicity we will assume the theory possesses a symmetry group $G_A\times G_B$. Gauging a symmetry group $G_A$ amounts to chopping up the space-time manifold into separate pieces, and stitching them back together up to a symmetry transformation in $G_A$. This procedure can also be thought of as endowing a system with twisted boundary conditions in the group $G_A$ in various directions. If, in the presence  of such twist, no local formulation of the theory is invariant under the $G_B$ group, then it must be that the boundary twists in the group  $G_A$ have excited a state in the group $G_B$. This can only happen if the ground state is nontrivial. In this case, we say that there exists a mixed 't Hooft anomaly between the global symmetry group $G_A$ and $G_B$. 

In order to gauge symmetries with 't~Hooft anomalies, 
nonlocal terms in the action are needed by definition. A typical way to achieve this is by extending the system into an auxiliary ``bulk'' direction which is in the symmetry-protected topological (SPT) phase. The anomaly inflow~\cite{Callan:1984sa} from the bulk SPT phase shows that the 't Hooft anomaly is renormalization-group invariant. 
The consequence of this anomaly matching condition is that the vacuum cannot be trivially gapped, and must either have a long-range order (i.e. spontaneous symmetry breaking), long-range entanglement (i.e. topological order) or infinite correlation length (i.e. gapless excitations)~\cite{Vishwanath:2012tq, Wen:2013oza, Kapustin:2014lwa, Kapustin:2014zva, Cho:2014jfa}. We emphasize that the auxiliary SPT bulk phase is not necessary to realize ``the boundary''  system, and is a priori a mathematical tool. Nevertheless a bulk may exist, and systems which are 't~Hooft anomalous may be realized as a boundary of some SPT bulk state. 

In previous studies~\cite{Gaiotto:2017yup,Komargodski:2017dmc, Komargodski:2017smk}, it was shown that the $\mathbb{C}P^{N-1}$ sigma model at $\theta=\pi$ has an 't~Hooft anomaly involving two-form gauge field between symmetries $SU(N)/\mathbb{Z}_N$ and charge conjugation $C$. 
For $N=2$, it is the $O(3)$ nonlinear sigma model and reproduces the Haldane conjecture for half-integer spins, so this 't~Hooft anomaly matching corresponds to the conventional LSM theorem requiring the spin-$SO(3)$ (or rather it's $O(2)$ subgroup) and  lattice translational symmetries. 
Similar 't~Hooft anomalies involving discrete symmetries have also been of recent interest in various gauge theories~\cite{Gaiotto:2017yup,Tanizaki:2017bam, Shimizu:2017asf, Gaiotto:2017tne, Tanizaki:2017qhf, Tanizaki:2017mtm, Cherman:2017dwt, Guo:2017xex}. 

In this paper, we show that a certain class 1+1D systems with a topological term 
have another kind of 't~Hooft anomaly which does not require the $SO(3)$ spin symmetry to be preserved at all. More concretely, we show that the $O(3)$ nonlinear sigma model at $\theta=\pi$ has a mixed 't~Hooft anomaly for the time-reversal $T$, spatial reflection $P$, and charge conjugation $C$ symmetries. We call this 't Hooft anomaly as a \CxPxT anomaly. Ref.~\cite{Metlitski:2017fmd} also discussed\footnote{We thank Z. Komargodski for pointing out this reference to us, which motivated our current work. } anomalies which do not involve the $SO(3)$ spin symmetry. However these are claimed to be IR emergent anomalies, involving spatial translations only. In contrast the anomaly we discuss here are fundamental anomalies of the underlying system, of the LSM-type, and they involve time-reversal and parity as well as spatial translations. 
The implication therefore is that the \CxPxT  anomaly is a different kind of  LSM theorem for spin chains which preserve parity, time reversal, and lattice translational symmetries~\cite{chen2011, PhysRevB.93.104425}, but not necessarily the $SO(3)$ spin symmetry. 

To show its existence, we put the theory on the non-orientable manifold $\mathbb{R}P^2$, along with certain boundary conditions on the fields, which amounts to gauging the $T$ and $CP$ symmetry (see also~\cite{Kapustin:2014tfa, Cho:2015ega, Hsieh:2015xaa,Metlitski:2015yqa, Witten:2016cio, Tachikawa:2016cha, Tachikawa:2016nmo, Tiwari:2017wqf} for related works). 
We find that the topological charge of the $O(3)$ sigma model becomes quantized to half integers on $\mathbb{R}P^2$, and the charge conjugation symmetry at $\theta=\pi$ is explicitly broken. Furthermore we show that there is no local counterterm to eliminate this $C$ breaking. 
In addition we generalize the \CxPxT  anomaly to the case of $\mathbb{C}P^{N-1}$ model when $N$ is even. 

We check that the constraint by the anomaly matching is consistent with the known results about our model and what is known about the corresponding spin systems, such as the XXZ and the XYZ model which are exactly solvable. 
We also discuss consistency of the anomaly with the semi-classical regimes. 
This will be helpful for clear understanding of how the anomaly matching is realized in this system. Finally we briefly discuss the 't Hooft anomaly of the underlying lattice system.

The paper is organized as follows: In Sec.~\ref{sec:O(3)_sigma}, we explain the $O(3)$ nonlinear sigma model and its discrete symmetries $C$, $P^*$, $T$. In Sec.~\ref{sec:CPT_anomaly}, we compute the 't~Hooft anomaly for $C$, $P^*$, and $T$ symmetries, and discuss semi-classical limits of the system. 
In Sec.~\ref{sec:generalization}, we discuss the generalization of the \CxPxT anomaly for the $\mathbb{C}P^{N-1}$ model for even $N$. In Sec.~\ref{sec:lattice} we discuss the anomaly in the underlying lattice system.
We give conclusions in Sec.~\ref{sec:conclusion}. 
In Appendix~\ref{app:anomaly_matching}, we give a brief review on anomaly matching. 

\section{$O(3)$ nonlinear sigma model}\label{sec:O(3)_sigma}

We will consider the $O(3)$ (i.e. $S^2$ target space) nonlinear sigma model in two-dimension with a $\theta$ term. The Lagrangian is 
\be\label{eq:O3model}
\mathcal L= |d\bm n|^2+\frac{i\theta}{8\pi}\bm n\cdot(d\bm n\times d\bm n), 
\ee
where $\bm n=(n_1,n_2,n_3)$ is the unit vector, $\bm n^2=1$. The second term ($\equiv i\theta Q$) is the topological theta term of the action, where $Q\in \Z$ on a closed orientable manifold. As a consequence, the partition function is periodic with respect to $\theta\rightarrow \theta+2\pi$.

We will first focus on model \eqref{eq:O3model}. However it is worth noting that everything we say will hold also for 
\be\label{eq:AH}
|Du|^2+m^2 |u|^2+V(u)+\frac{1}{4e^2}F\wedge \star F+\frac{i \theta}{2\pi}F,
\ee
where $u=(u_1,u_2)$ is a scalar doublet, $D=d+iA$ is the $U(1)$ gauge-covariant derivative, $F=dA$ is the field strength, $V(u)$ is a gauge invariant potential of $u$ obeying the relevant symmetries. The two models are equivalent if we set $u^\dagger u=1, V(u)=0$ and the $F^2$ term is dropped. 

The system \eqref{eq:O3model} has a time-reversal\footnote{How time-reversal -- an anti-unitary symmetry -- acts in the Euclidean path integral is slightly subtle and is addressed in the appendix~\ref{app:Noether}. The result is simple enough, as it amounts to the reversal in time in the Euclidean Lagrangian (without complex conjugation), without affecting the limits of the time integration.} and parity (or equivalently -- reflection) symmetry
\begin{subequations}\label{eq:TRsymmetry}
\begin{align}
T(\bm n(x,t))=-\bm n(x,-t)\\
P^*(\bm n(x,t))=-\bm n(-x,t)
\end{align}
\end{subequations}
at any $\theta$. Notice that we use the asterisk on $P$. This is because this transformation has two interpretations: either as a $CP$ symmetry of the field theory or a bond-centered parity symmetry of the underlying spin-chain. 

\begin{figure}[t] 
   \centering
   \includegraphics[width=3.4in]{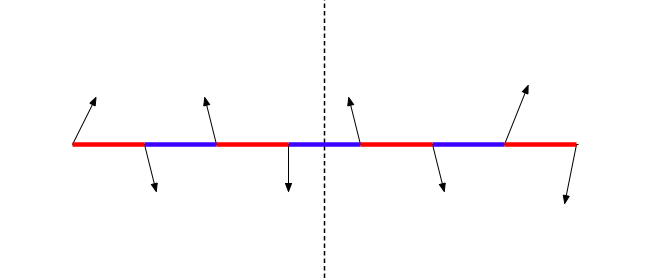} 
   \caption{A cartoon of a spin chain without $\bmod 2$ translational symmetry. The red/blue links indicate that the interactions are different. The dashed line indicates the axis around which parity is a good symmetry.}
   \label{fig:spin-chain}
\end{figure}

Let us explain this a bit more. The above theory is believed to be an effective theory of an anti-ferromagnetic spin-chain. The chain has a translational symmetry which is a symmetry of the Hamiltonian. The continuum limit then requires that spin $\bm s_i$ at site $i$ is related to the N\'eel vector $\bm n_i$ at the same site by $\bm n_i=(-1)^i \bm s_i$, i.e. the N\'eel vector is staggered by a different sign on the two sublattices. The translation symmetry by a single lattice unit therefore changes the sign of the N\'eel vector. So in the field theory description \eqref{eq:O3model}, this $\mathbb Z_2$ translational symmetry ($\bmod$ $2$) is mapped to the charge conjugation symmetry 
\be
C(\bm n(x,t))=-\bm n(x,t).
\ee 
and it is only a symmetry of \eqref{eq:O3model} when $\theta=0,\pi$. Hence the arbitrary $\theta$ corresponds to the absence of the $\bmod$ $2$ translational symmetry. Microscopically the lack of $\mathbb Z_2$ translational symmetry is sketched in Fig.~\ref{fig:spin-chain}, where red and blue bonds denote different anti-ferromagnetic interactions. Notice that the parity around the midpoint of any link is a good symmetry. This parity acts by interchanging the two sub-lattices, which has the effect of changing the N\'eel order parameter
$\bm n\rightarrow -\bm n$. So the operator $P^*$ is a link-centered parity of the underlying microscopic theory\footnote{The situation is reminiscent to identifying a field-theory $CT$ symmetry with the microscopic $T$-symmetry as in Ref.~\cite{Seiberg:2016rsg}.}.

On the other hand  $P^*$ could also correspond to the vertex centered reflection followed by a $\mathbb Z_2$ translation symmetry as well. Since the $\mathbb Z_2$ translation symmetry is mapped to the $C$-symmetry in field theory we would call the combination of the two \emph{the $CP$ symmetry}.

In fact the model \eqref{eq:O3model} is equivalent to the $\mathbb{C}P^1$ model, where instead of $\bm n$ fields we write the action in terms of the $u=(u_1,u_2)$ --- a complex $SU(2)$ doublet, and $A=-i u^\dagger d u$ --- the auxiliary gauge field. It can be checked (see below) that the $P^*$ symmetry acts on the theory as the $CP$ symmetry, where $C$ is charge conjugation. 
Indeed, the $O(3)$ sigma model (\ref{eq:O3model}) has the $CPT$ invariance: We have $\bm n(x,t)\to \bm n(-x,-t)$, by Lorentz invariance, and $P^*$ in (\ref{eq:TRsymmetry}) is the $CPT$ partner of the time-reversal symmetry $T$. 

\section{Gauging $P^*$ and $T$ symmetries}\label{sec:CPT_anomaly}

We will now show that at $\theta=\pi$ there exists a mixed 't~Hooft anomaly between $T, P$ and $C$ symmetry. 
To see this, let us first place the system on a $2$-torus, so that the coordinate $x\in[-L/2,L/2]$ and $t\in[-\beta/2,\beta/2]$. The fields obey boundary conditions
\be
\bm n(x+n L,t+m\beta)=\bm n(x,t), \quad \forall n,m\in \Z\;.
\ee
The path integral over all the possible configurations with this boundary condition gives the ordinary partition function of the system.

Now we wish to gauge the $P^*$ and $T$ symmetries \eqref{eq:TRsymmetry} in order to detect the 't Hooft anomaly involving them. To that end, we put the theory on a non-orientable manifold by imposing the boundary conditions
\begin{subequations}\label{eq:nBC}
\begin{align}
&\bm n(L/2,t)=T(\bm n(-L/2,t))=-\bm n(-L/2,-t),\\
&\bm n(x,\beta/2)=P^*(\bm n(x,-\beta/2))=-\bm n(-x,-\beta/2)\;.
\end{align}
\end{subequations}
We interpret these boundary conditions as putting the background gauge fields for the symmetries \eqref{eq:TRsymmetry} along the two cycles of the torus\footnote{The twist in the in the temporal direction can be interpreted as an insertion of a parity symmetry operator into the thermal partition function. This can be interpreted as the presence of a nontrivial parity $\mathbb Z_2$ gauge field in the temporal direction. Similarly spatial gauge fields were constructed in Ref.~\cite{Thorngren2016}, to generalize the notion of gauge fields in the spatial directions. We apply the same logic to the time-reversal symmetry, and implement time reversal at some point along the spatial circle. However we here emphasize that since the time-reversal symmetry is an antiunitary symmetry, we do not know whether a Hilbert space interpretation of twists exists, i.e. whether it makes sense to gauge the $T$-symmetry with a temporal gauge field. Part of the problem is that there is a difficulty of defining the $T$-operator insertion into the partition function, because trace of the antilinear operator is not invariant under the basis change. This is why we circumvent this problem by applying the gauge field of the $T$ symmetry along the spatial direction.}. 
Quick thought reveals that the boundary conditions render the manifold topologically equivalent to a disc $D$ with any two opposing points on its boundary identified. 
This manifold is a projective plane $\mathbb{R}P^2$ and is non-orientible. 

\subsection{Topological charge on $\mathbb{R}P^2$}\label{sec:top_charge}

\begin{figure}[t] 
   \centering
   \includegraphics[width=3.4in]{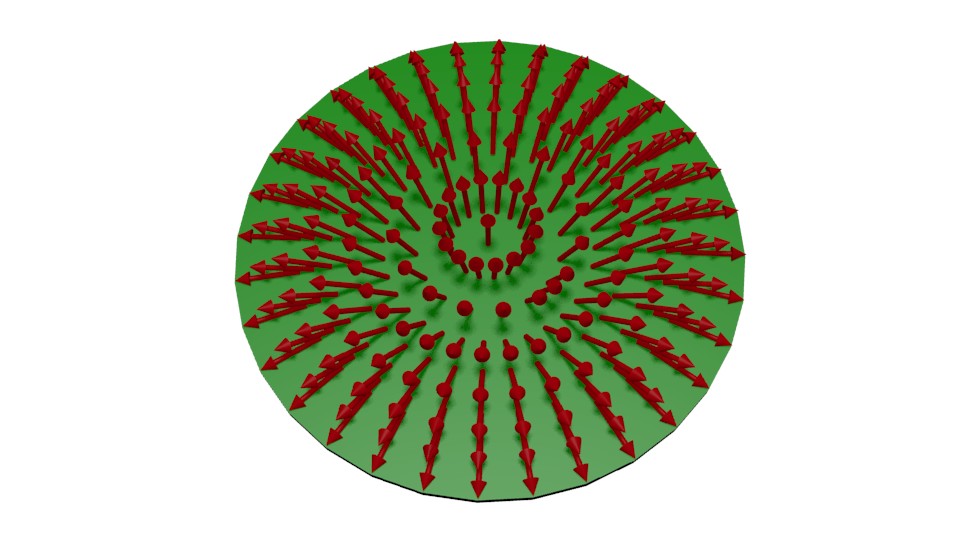} 
   \caption{A figure of a half-instanton configuration on a disk $\mathbb RP^2$, which is topologically a disk whose opposite points on the edges are identified up to a transformation $\bm n\rightarrow -\bm n$. The red arrows represent the unit vector $\bm n$. It is clear that the configuration depicted has a half-integer winding number.}
   \label{fig:half_instanton}
\end{figure}

We wish to show that with these boundary conditions (\ref{eq:nBC}) the quantization of topological charge is in half-integer units. Namely the topological charge is $Q=1/2\bmod 1$. In fact we can see this already by considering $\mathbb{R}P^2$ to be topologically a disk, with opposing points on the boundary of the disc identified up to a transformation $\bm n\rightarrow -\bm n$. 
An example of such a configuration is given by $\bm n=(\sin(\theta)\cos(\phi),\sin(\theta)\sin(\phi),\cos(\theta))$, where we parametrize the disk $D$ by the stereographic projection $(x,y)=(\sin(\theta)\cos(\phi),\sin(\theta)\sin(\phi))$ for $0\le \theta\le \pi/2$ and $-\pi\le \phi\le \pi$ (see Fig.~\ref{fig:half_instanton}). The topological charge is 
\bea
Q&=&{1\over 8\pi}\int_{0}^{\pi/2}\diff \theta\int_{-\pi}^{\pi}\diff \phi \, 2 \bm n \cdot (\p_{\theta} \bm n\times \p_{\phi}\bm n)\nonumber\\
&=&{1\over 4\pi}\int_{0}^{\pi/2}\diff \theta\int_{-\pi}^{\pi}\diff \phi \sin(\theta)
={1\over 2}. 
\eea
This configuration indeed has a half-integer winding number. This configuration is sketched in Fig.~\ref{fig:half_instanton}, and it obviously covers only the half of the target space $S^2$. 

Let us now prove that all configurations have a half-integral winding number. To do this it is useful to use the $\mathbb{C}P^1$ parametrization, i.e. to define
\be
\bm n=u^\dagger \bm \tau u\;, 
\ee
where $u\in \mathbb{C}^2$ with the constraint $u^\dagger u=1$. 
Here, $\tau$ is the Pauli matrix, $[\tau^a,\tau^b]=2\im \ve^{abc}\tau^c$. 
This parametrization has a $U(1)$ gauge symmetry $u\rightarrow e^{i\phi}u$. Then it can be shown that
\be
\frac{i}{4}\int \bm n\cdot(d\bm n\times d\bm n)=\int du^\dagger \wedge du= i\int dA
\ee
where we defined 
\be
A=-iu^\dagger d u\;.
\ee
This is a $U(1)$ gauge field.

Note that in this parametrization, the transformation $\bm n\rightarrow -\bm n$ corresponds to $u\rightarrow i\tau^2 u^*$, where $^*$ denotes complex conjugation. Therefore we have that $u$ and $A$ transform as
\begin{align}
&T(u(x,t))=i\tau^2u^*(x,-t),\\
&P^*(u(x,t))=i\tau^2 u^*(-x,t)\;,
\end{align}
while 
\begin{align}
&T(A_x(x,t))=-A_x(x,-t),\\
&T(A_t(x,t))=A_t(x,-t),\\
&P^*(A_x(x,t))=A_x(-x,t),\\
&P^*(A_t(x,t))=-A_t(-x,t)\;.
\end{align}
The $T$ transformation is the usual transformation of the gauge fields under time-reversal\footnote{An easy way to see this is that the coupling of the gauge field to the conserved current  $A_\mu j^\mu$ must be invariant under the $T$ transformation. Now under the $T$-transformation the charge density is invariant $\rho_t$, however the spatial current $j_x$ flips sign under $T$.}. On the other hand the $P^*$ transformation is not what we usually call parity transformations. Instead in this context  it is identified with $CP$ (see discussion below \eqref{eq:TRsymmetry}). 

Once we have established the transformations, we can shift our attention entirely to the gauged linear sigma model, of the type \eqref{eq:AH}, and everything we say below is equally applicable to this case.

\begin{figure}[t] 
   \centering
   \includegraphics[width=3.4in]{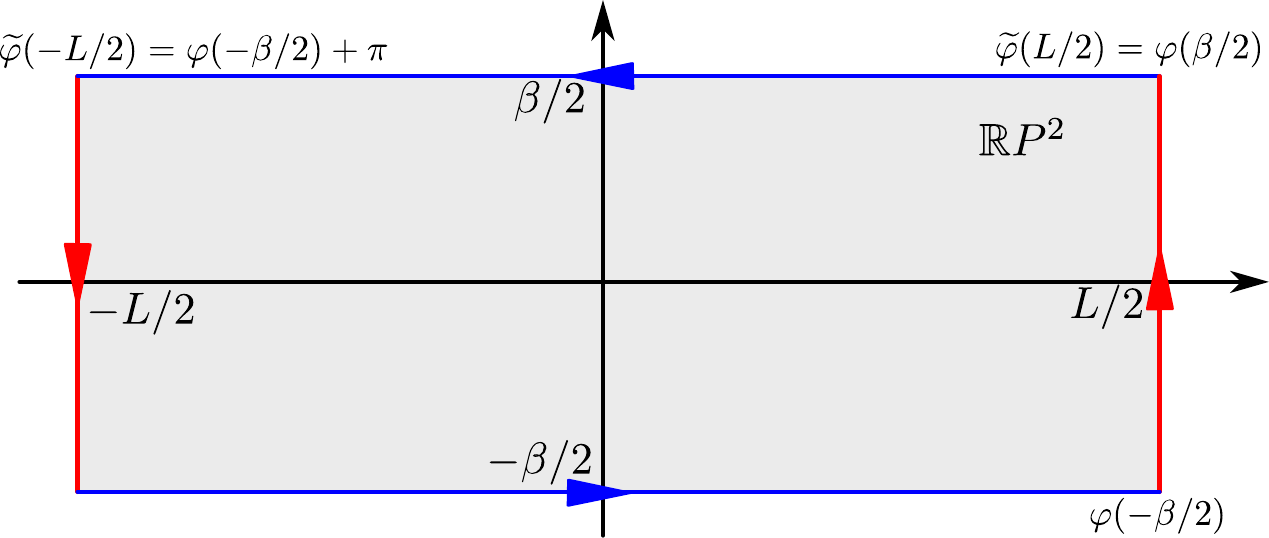} 
   \caption{A figure illustrating the twisted boundary condition for reflection symmetries with the $\mathbb{C}P^1$ representation.}
   \label{fig:CP1_boundary}
\end{figure}

The boundary conditions \eqref{eq:nBC} turn into
\begin{subequations}\label{eq:uBC}
\begin{align}
&u(L/2,t)=ie^{i\varphi(t)}\tau^2 u^*(-L/2,-t)\\
&u(x,\beta/2)=ie^{i\tilde\varphi(x)}\tau^2 u^*(-x,-\beta/2)\;.
\end{align}
\end{subequations}
and
\begin{subequations}\label{eq:ABC}
\begin{align}
&A_t(L/2,t)=A_t(-L/2,-t)+\partial_t\varphi(t)\\
&A_x(L/2,t)=-A_x(-L/2,-t)\\
&A_t(x,\beta/2)=-A_t(-x,-\beta/2)\\
&A_x(x,\beta/2)=A_x(-x,-\beta/2)+\partial_x\tilde\varphi(x)
\end{align}
\end{subequations}
By plugging $t=-\beta/2$ into the first equation of \eqref{eq:uBC}, and $x=-L/2$ into the second equation, we get

\begin{align*}
&u(L/2,-\beta/2)=ie^{i\varphi(-\beta/2)}\tau^2 u^*(-L/2,\beta/2)\\
&u(-L/2,\beta/2)=ie^{i\tilde\varphi(-L/2)}\tau^2 u^*(L/2,-\beta/2)\;.
\end{align*}
Complex conjugating the second equation, and plugging into the first, we get
\be
u(L/2,-\beta/2)=e^{i(\varphi(-\beta/2))-i\tilde\varphi(-L/2)+\pi)}u(L/2,-\beta/2)\;,
\ee
hence we conclude that
\be
\varphi(-\beta/2)-\tilde\varphi(-L/2)=\pi\bmod 2\pi\;.
\ee
Similarly, by putting $t=\beta/2$ in the first, and $x=L/2$ in the second equation of \eqref{eq:uBC} we get
\be
\varphi(\beta/2)=\tilde\varphi(L/2)\bmod 2\pi\;.
\ee
The result is summarized in Fig.~\ref{fig:CP1_boundary}. 

Now we are ready to evaluate the $\theta$-term on the $\mathbb RP^2$ manifold which arose from gauging the $P^*$ and $T$ symmetries. For this purpose we will consider $RP^2$ to be a filled rectangle $R$ depicted in Fig.~\ref{fig:CP1_boundary}, where horizontal axis represents space and vertical axis represents Euclidean time. The upper and lower edges of the rectangle are identified modulo the $P^*$ transformation which reverses space, while the left and right edges are identified modulo $T$-transformation which reverses time, which is indicated by the arrows of the respective edges.  We write
\be
\int_R dA=\int_{\partial R} A, 
\ee
where the integral on the right is along the boundary of our open rectangle $R$. 
Now we obtain that
\bea
\int_{R}d A&=&
\int_{-L/2}^{L/2} (A_x(-x,-\beta/2)-A_x(x,\beta/2))dx\nonumber\\
&&+\int_{-\beta/2}^{\beta/2}(A_t(L/2,t)-A_t(-L/2,-t))dt\nonumber\\
&=&-\int_{-L/2}^{L/2} \partial_x\tilde\varphi(x)dx+\int_{-\beta/2}^{\beta/2}\partial_t\varphi(t)dt\nonumber\\
&=&-\tilde\varphi(L/2)+\tilde\varphi(-L/2)+\varphi(\beta/2)-\varphi(-\beta/2)\nonumber\\
&=&\pi \qquad (\bmod 2\pi). 
\label{eq:mixed_anomaly_CPT_1}
\eea
where we identified the appropriate $P^*$ and $T$ relations on the upper/lower and left/right edges of Fig.~\ref{fig:CP1_boundary} respectively.
This shows that all the configurations with boundary conditions \eqref{eq:nBC} have a half integral winding number.

\subsection{Non-periodicity in $\theta$ and the mixed 't Hooft anomaly between $T, P$ and $C$}

As we already discussed the model \eqref{eq:O3model} has $T$ and $P^*(=CP)$ symmetries for any value of $\theta$. We have showed above how we can put background gauge fields for the $T$ and $P^*$ symmetry. 
However the model also has another symmetry $\bm n(x,t)\rightarrow -\bm n(x,t)$ when $\theta=0,\pi \bmod 2\pi$, i.e. the charge conjugation $C$. 
When $\theta=\pi$, the $C$-symmetry is a symmetry because the winding number is quantized in integer units. Similarly the partition function is a periodic function of $\theta$ because of this quantization.

However as we already saw when we gauge the $T$ and $P^*$ symmetries, the quantization of topological charge is in half-integer units, and the periodicity of the partition function with respect to $\theta\rightarrow \theta+2\pi$ is ruined. For $\theta=\pi$ this has a consequence that the $C$ symmetry is not a symmetry in the presence of the $T$ and $P^*$ gauge fields. Equation~(\ref{eq:mixed_anomaly_CPT_1}) shows that the partition function on the non-orientable manifold $\mathbb{R}P^2$, $\mathcal{Z}_{\mathbb{R}P^2}$, is changed under $C$ as 
\be
\mathcal{Z}_{\mathbb{R}P^2}\to (-1)^{2S} \mathcal{Z}_{\mathbb{R}P^2}
\label{eq:mixed_anomaly_CPT_2}
\ee
with $\theta=2\pi S$, and the $C$ symmetry is broken while it is a good symmetry on any orientable manifolds. 

We interpret this as the mixed 't Hooft anomaly between $T$, $P^*$, and $C$ at $\theta=\pi$. For that purpose, we must check that the anomaly (\ref{eq:mixed_anomaly_CPT_2}) is not fake, i.e. it cannot be canceled by the local counterterm to the action. 
To show it, we have to look at all the possible $(1+1)$D local actions as we shall do below. 

In fact, the possible local counterterm is characterized by the topological action depending on the gauge fields for $T$, $P^*$, and it is given by 
\be
S_{\mathrm{counter}}={i\pi} \int_{\mathbb{R}P^2} w_1(E)^2\;. \label{eq:anomaly_polynomial}
\ee
Here, $E$ is the real-vector bundle on $\mathbb{R}P^2$ whose section defines the real-vector field $\bm n$, and $w_1(E)\in H^1(\mathbb{R}P^2,\mathbb{Z}_2)$ is the first Stiefel-Whitney class, which is the obstruction of orientation.

Let us explain the notation a bit. We can consider $w_1$ to be the $\mathbb Z_2$ gauge field for the $T$ and $P^*$ symmetry. So if the integral of $w_1$ over some 1-cycle is nontrivial, it means that a transformation reversing the orientation of space-time and flipping $\bm n\rightarrow -\bm n$ is applied as one traverses the cycle. Further, notice that what we have shown in the previous section is that
\be
e^{i\int F}=e^{i\pi \int w_1^2}\;.
\ee
The right hand side is trivial on an orientable manifold, as it should be.
 
The second Stiefel-Whitney class $w_2(E)\in H^2(\mathbb{R}P^2,\mathbb{Z}_2)$ is also a candidate for the counterterm. 
On $2$-dimensional manifolds, however, there is a relation $w_2(E)=w_1(E)^2(\equiv w_1(E)\cup w_1(E))$ due to vanishing of the second Wu class, and thus this topological action (\ref{eq:anomaly_polynomial}) is the unique possible counterterm (see, e.g., Appendix of Ref.~\cite{Kapustin:2014gma}). 
Here, $\cup$ is the cup product.
The above integral does not vanish on the non-orientable manifold, while it always vanishes on orientable manifolds modulo $2\pi$. 
The charge conjugation does not change the counterterm $S_{\mathrm{counter}}$ at all, and thus it cannot eliminate the anomaly (\ref{eq:mixed_anomaly_CPT_2}). 

Finally notice that if we continuously deform a theory by changing its $\theta$ term by $2\pi$, we generate a counterterm $S_{\mathrm{counter}}$. If the theory is defined on a manifold with boundaries, the partition function is supplemented with a factor $e^{S_{\mathrm{counter}}}$, which requires that the boundaries are carrying Kramers' doublets~\cite{Kapustin:2014gma}.

We therefore have an 't~Hooft anomaly between $T$, $P$ and $C$ symmetries at $\theta=\pi$. The anomaly matching claims that the system must either
\begin{itemize}
\item break $C$, $P$, or $T$ symmetry spontaneously\footnote{If we assume the theory to be relativistic, and if we assume that the ground state does not break Lorentz symmetry spontanously, we must have that if $T$ is broken, then $P^*(=CP)$  must also be broken.},
\item have massless excitations (conformal behavior), or
\item 
have long-range entanglement (topological order).
\end{itemize}
For 1D spin chains, it is shown that the topological order does not appear~\cite{chen2011}. Therefore, we can conclude that the system is conformal or breaks one of the discrete symmetries if the field theory appears as a low-energy description of the spin model. 
When the field theory does not have the UV Hamiltonian of the 1D spin systems, however, this additional constraint does not apply. In fact we will show that a topological phase can arise as a limit, in section Sec.~\ref{sec:checks}. This scenario is quite esoteric however, and arises only if the charge-2 Higgs field has a runaway potential. Since we are unaware of of a scenario which would contain a charge-2 condensing Higgs field\footnote{It may be possible to obtain such exotic phases by realizing 1D systems as domain walls of higher dimensional systems, as in Refs.~\cite{Sulejmanpasic:2016uwq, Komargodski:2017smk} (see also Ref.~\cite{Wang:2018edf})}, let alone one with a runaway potential, we will exclude this possibility from our considerations. 

We can make a stronger conclusion by taking into account the spin rotation symmetry $SO(3)$. In Refs.~\cite{Komargodski:2017dmc, Komargodski:2017smk}, it is shown that there is a mixed 't Hooft anomaly between $SO(3)$ and $C$, which derives the conventional LSM theorem. Therefore, we must match both the $SO(3)$-$C$ and \CxPxT anomalies for the $O(3)$ sigma model at $\theta=\pi$.  
This means that, for instance, the spontaneous time-reversal breaking is not enough to match the anomaly, because it matches the \CxPxT anomaly but does not match the $SO(3)$-$C$ anomaly. 
By taking into account the Coleman-Mermin-Wagner theorem~\cite{Coleman:1973ci, mermin1966absence}, the $SO(3)$ symmetry cannot be broken spontaneously in two-dimension, and the anomaly is matched by 
\begin{itemize}
\item spontaneous breaking of $C$, or
\item conformal behavior
\end{itemize}
It is also interesting to consider the situation where $SO(3)$ symmetry is broken down to some discrete subgroup $K$ explicitly by additional interactions, and let us assume that there is the mixed anomaly between $K$ and $C$ as a remnant of the $SO(3)$-$C$ anomaly.  The $K$-$C$ anomaly and the \CxPxT anomalies are matched by 
\begin{itemize}
\item spontaneous breaking of $C$,
\item spontaneous breaking of $K$, and of $P$ or $T$, or
\item conformal behavior
\end{itemize}

\subsection{Consistency checks of the anomaly}\label{sec:checks}

\subsubsection{Consistency with the XYZ model}

We will now comment on the XYZ model which is exactly solvable, and see how its phases match the anomaly. This will provide more insight about anomaly matching for strongly-coupled cases. The Hamiltonian is given by
\be
H_{\mathrm{XYZ}}=-\sum_{i=1}^{L}(J_x \tau_{x,i}\tau_{x,i+1}+J_y \tau_{y,i}\tau_{y,i+1}+J_z\tau_{z,i}\tau_{z,i+1}),
\ee
with the periodic boundary condition, $\bm \tau_{L+1}=\bm \tau_{1}$, and this system is exactly solvable by the Bethe ansatz (see, e.g., Refs.~\cite{baxter2007exactly, takahashi1999thermodynamics}). 
This Hamiltonian still have a discrete $K_4=\mathbb Z_2\times \mathbb Z_2$ subgroup\footnote{$K_4$ is the Klein four group and is generated by $\pi$ rotations of $SO(3)$ around any two axis.} of $SO(3)$ symmetry, so it is not completely generic.
The model also possess, in addition to the spin $K_4$ symmetry, the translational symmetry which is mapped to $C$ in the field theory language, the time-reversal symmetry $T$ and the bond-centered parity $P^*(=CP)$. Further in addition to the $\CxPxT$ mixed 't Hooft anomaly anomaly, it also possesses the mixed anomaly between\footnote{See \cite{Jian:2017skd} and the Appendix \ref{app:K4-C}.} $K_4$ and $C$.

 Still, it is insightful to compare our anomaly with its ground-state structures. 
When $J_x=J_y\equiv J$, this is called the XXZ model. It preserves $O(2)\subset SO(3)$ and its ground-state properties are parametrized with $\Delta=J_z/J$ as follows: 
For $\Delta>1$, the ground state is doubly degenerate, which is the ferromagnetic phase\footnote{Those ground states are given by  $|\up\up\up\ldots\up\rangle$ and $|\down\down\down\ldots\down\rangle$.~\cite{takahashi1999thermodynamics}}. The anomaly is matched by spontaneous $T$ breaking. It also breaks the $K_4$ down to $\mathbb Z_2$, and the spontaneous breaking of $K_4$ and $T$ is required to match both $K_4$-$C$ anomaly and \CxPxT anomaly\footnote{Note that the effective theory of the ferromagnets is not described by the relativistic theory in the IR (see e.g.~\cite{Altland:2006si}) because of the spin Berry phases. Hence breaking $T$ without breaking $P^*$ is allowed.}. 
For $\Delta<-1$, the ground state is nearly degenerate at finite $L$, and the difference between them vanishes as $O(1/L)$ in the thermodynamic limit $L\to \infty$ \cite{takahashi1999thermodynamics}. This is the antiferromagnetic phase\footnote{In the antiferromagnetic Ising limit, $\Delta\ll -1$, those two states are given by $|\up\down\up\down\cdots\up\down\rangle\pm |\down\up\down\up\ldots\down\up\rangle$ for even $L$}, which breaks the $\bmod$ $2$ translational symmetry spontaneously (and $T$ symmetry) in the thermodynamic limit. Therefore, the anomaly is matched by $C$ and $T$ breaking. 
When $|\Delta|<1$, we have gapless excitations~\cite{takahashi1999thermodynamics}, called spinon, spin-wave and bound-state excitations, and they match the anomaly. 
For the most general XYZ model, it is sufficient to consider the case $J_z> J_y\ge |J_x|\ge 0$ since it is known that the spectrum of the XYZ Hamiltonian is invariant under the flip of two $J$'s. 
This system is gapped, but we can construct the nearly degenerate state, whose energy difference from the ground-state energy rapidly decreases in the thermodynamics limit $L\to \infty$ \cite{takahashi1999thermodynamics}. Therefore, the anomaly is matched by the spontaneous symmetry breaking. Depending on the parameters $J_x,J_y$ and $J_z$ (all which are different), the system either is in a ferromagnetic or anti-ferromagnetic phase. The earlier phase breaks the $T$-symmetry as well as $K_4$ and the latter breaks $T$ and translational symmetry. 

\subsubsection{Consistency with semi-classics}

Let us now explore some semi-classical limits. It is believed that the model \eqref{eq:O3model} at $\theta=\pi$ is described by the $SU(2)_1$ Wess-Zumino-Witten conformal theory. This is consistent with the previously found 't Hooft anomaly in the $O(3)$ sigma model between $SO(3)$ and $C$ symmetries~\cite{Komargodski:2017dmc, Komargodski:2017smk}.

As we have seen, however, that 't Hooft anomaly exists even if the $O(3)$ global symmetry is reduced down to its improper $\mathbb Z_2$ part, as long as parity and time-reversal are good symmetries of the theory. We will now explore deformations which will keep the \CxPxT anomaly intact, and demonstrate several regimes how the anomaly manifests itself. Thus, we focus on an (Euclidean) action
\be
|Du|^2+V(u)+\frac{1}{4e^2}F\wedge \star F+\frac{i \theta}{2\pi}F,
\ee
where $u=(u_1,u_2)$ -- a complex SU(2) doublet, $F=dA$ is the $U(1)$ gauge field strength of the dynamical $U(1)$ gauge field $A$, and  $D=d+i A$ is the covariant derivative. $V(u)$ is a gauge invariant potential for the $u$-scalars, which also preserves the $C$-symmetry defined as
\be
u\rightarrow \im \tau^2 u^*\;,\\
A\rightarrow -A\;.
\ee

For simplicity of notations to construct such $V(u)$, let us define the gauge-singlet fields
\be
n_0=u^\dagger u,\;\; \bm n=u^\dagger \bm\tau u=(n_1,n_2,n_3)\;. 
\ee
$n_0$ and $\bm n$ are related by $n_0=\sqrt{\bm n^2}$, but we do not require that $\bm n^2(=n_0^2)$ is unity. The $C$ transformation changes the sign of all the components of $\bm n$, while $n_0$ does not change under $C$. Hence, to preserve the C-symmetry, we must have that the potential is an arbitrary function of $n_0$ and of the product $n_i n_j$ for any $i,j=1,2,3$, i.e. 
\be
V(u)=V(n_0,\{n_in_j\}_{i,j=1,\ldots,3})\;.
\ee 
Such a potential in generic case completely destroys the $SO(3)$ symmetry. Yet, our consideration guarantees that the ground state cannot be trivially gapped.

Classically speaking, if the potential $V(n^0,\{n_in_j\})$ is minimized at any nonzero value of $\bm n$ it must break the $C$-symmetry, and have at least two classically degenerate vacua. On the other hand, if a potential minimizes at $\bm n=0$ we must have $u=0$. Then, semi-classically speaking, we can integrate out the massive $u$-fields, and obtain an effective pure-gauge theory (quantum Maxwell theory) at $\theta=\pi$. This theory again breaks $C$-symmetry at $\theta=\pi$, as is well known.  

In the end let us consider another,  purely quantum, scenario. Let us add a scalar with charge $q$ under the $U(1)$ gauge group, i.e. a field $\phi$ which has a kinetic term given 
\be
|(d+qiA)\phi|^2\;, q\in \mathbb Z\;.
\ee
Now if we wish to have the same $C$-$P^*$-$T$ anomaly, we must first check if the global symmetries are the same. To that end, consider a gauge-invariant operator
\be
O_q=\phi^* u_{\alpha_1}u_{\alpha_2}\dots u_{\alpha_q}\;.
\ee
Under the charge conjugation, $\phi$ is mapped to $\phi^*$, and $C^2(\phi)=\phi$. 
Now let us apply the $C$-transformation twice, then the above gauge-invariant operators are transformed as
\be
C^2:O_q\rightarrow (-1)^q O_q\;.
\ee
The $C$ symmetry is no longer a $\mathbb Z_2$ symmetry, but a $\mathbb Z_4$ symmetry if $q$ is an odd integer. Since we wish to consider possible deformations of the original model, without changing the global symmetries, we choose $q$ to be even.

Now the global symmetry group is unchanged, and the \CxPxT anomaly is present, so that we cannot have a trivially gapped state. Let us now take the mass-squared of the charge-$q$ scalar $\phi$ to be large and negative, so that it condenses and Higgses the gauge field. In other words we can write $\phi=ve^{-i\varphi}$, where we can ignore the fluctuation of $v$ and set it to be constant and large. Then the kinetic term gives
\be
v^2 |d\varphi-q A|^2\;.
\ee
Minimization of this term forces the $U(1)$ gauge field to be a $\mathbb Z_q$ gauge field:
\be
q A=\diff \varphi. 
\label{eq:Zq_gauge_field}
\ee
However, the theory can have gauge-vortices for which $\varphi$ winds by $2\pi$ as the nexus of the vortex is traversed, and the above condition (\ref{eq:Zq_gauge_field}) is satisfied only outside the vortex cores. The vortex therefore must have a fractional $1/q$ flux, because for the single-vortex
\be
\int_{\mathbb{R}^2} dA=\oint_{S^1_{\infty}} A={1\over q}\int_0^{2\pi}d\varphi=\frac{2\pi}{q}\;.
\ee
Because the vortices carry fractional flux, they can only occur in bunches of $q$ to satisfy the correct boundary condition for the thermal partition function. Furthermore, they will each couple to the $\theta$-angle as $e^{i{\theta}/{q}}$. If we naively sum over all such vortices by the dilute-instanton gas approximation (, i.e., ignoring their exponentially small interaction), we get that the energy density is given by
\be
\epsilon\sim -e^{-S_0}\cos\left(\frac{\theta}{q}\right)\;.
\ee
Here, $S_0\sim v$ is the action of the fractional vortex. The above expression is not periodic with respect to $q$. 
This is because we have not taken into account the fact that only bunches of $q$ vortices can contribute to the partition function. 
To achieve that, we must impose the constraint that $\int dA\in 2\pi \mathbb Z$ by introducing the Kronecker's delta, which it can be written as the Fourier series
\be
\sum_{k=0}^{q-1}e^{ik\int dA}\;.
\ee
Notice that the above integer combines with the $\theta$-angle, and amounts to replacing $\theta\rightarrow \theta+2\pi k$. Notice that the $C$-symmetry is now a symmetry at $\theta=\pi$ only if we take $k\rightarrow -k-1$. As a result, we obtain that the ground-state energy is
\be
\epsilon\sim \min_{k=0,\ldots,q-1}\left\{-e^{-S_0}\cos\left(\frac{\theta+2\pi k}{q}\right)\right\}\;.
\ee
At $\theta=\pi$ the vacua $k=0$ and $k=-1$ are degenerate. Since they are also related by the $C$ symmetry, we conclude that the system breaks $C$-symmetry.
%
%

Let us now take the limit of $v\rightarrow \infty$. In this case $C$-symmetry is restored since the fugacity of the fractional vortex vanishes, and we have an apparent contradiction with the anomaly. However notice that the $u_\alpha$ quanta are completely noninteracting now, and therefore become fractionalized excitations, so that there is still a long-range topological order. 
In other words, the constraint (\ref{eq:Zq_gauge_field}) holds true everywhere on the spacetime in this limit, and thus the low-energy field theory is described a nontrivial topological field theory, called $\mathbb{Z}_q$ BF theory. 

However, as we commented before, it is doubtful that such an instability is of physical relevance. In order to be relevant, one would have to obtain a form of a runaway potential for the Higgs field. We only mention such a phase as a curiosity. 

\section{Generalization of \CxPxT anomaly to $\mathbb{C}P^{N-1}$ model for even $N$.}
\label{sec:generalization}

In order to construct the appropriate $C$-symmetry, we want to define
\be
u\rightarrow \Omega u^*
\ee
where $\Omega$ is an $SU(N)$ matrix. If we apply the above transformation twice we get
\be
u\rightarrow \Omega \Omega^* u\;.
\ee
In order for the $C$-symmetry to be a $\mathbb Z_2$ symmetry we must require that $\Omega \Omega^*=z\bm 1\in \mathbb Z_N$ -- an element of the center. If this is the case with the nontrivial center, $z\not=1$, then the $C$-symmetry will act projectively and we may have an 't Hooft anomaly. 

The condition translates to
\be
\Omega=z^*\Omega^T
\ee
Since $z\in \mathbb{Z}_N\subset U(1)$, this is equivalent to
\be
\Omega=z\Omega^T\;
\ee
by taking the transpose of the both sides. 
In order for both to be true we must have that $\text{Im }z=0$.

For odd $N$ we have that the only element of the center with such a property is a trivial element $\bm 1$. So we conclude that there is no corresponding $C$-symmetry in the $\mathbb{C}P^{N-1}$ case for which there is a \CxPxT anomaly if $N$ is odd. 

For $N$-even we have that the element $z=-1$ is always in the center, hence it is possible to satisfy this condition. In fact the condition amounts to finding an anti-symmetric $SU(N)$ element.
Indeed if we pick
\be
\Omega=\text{diag}(\underbrace{i\tau^2,i\tau^2,\dots, i\tau^2}_{N/2\,\, \mbox{copies}})\;,
\ee
it is clear that this condition is satisfied. 
We can then follow the same steps as in Sec.~\ref{sec:top_charge} and show that there exists a \CxPxT anomaly.

\section{Comment on the microscopic spin chains}\label{sec:lattice}

\begin{figure}[htbp] 
   \centering
   \includegraphics[width=2in]{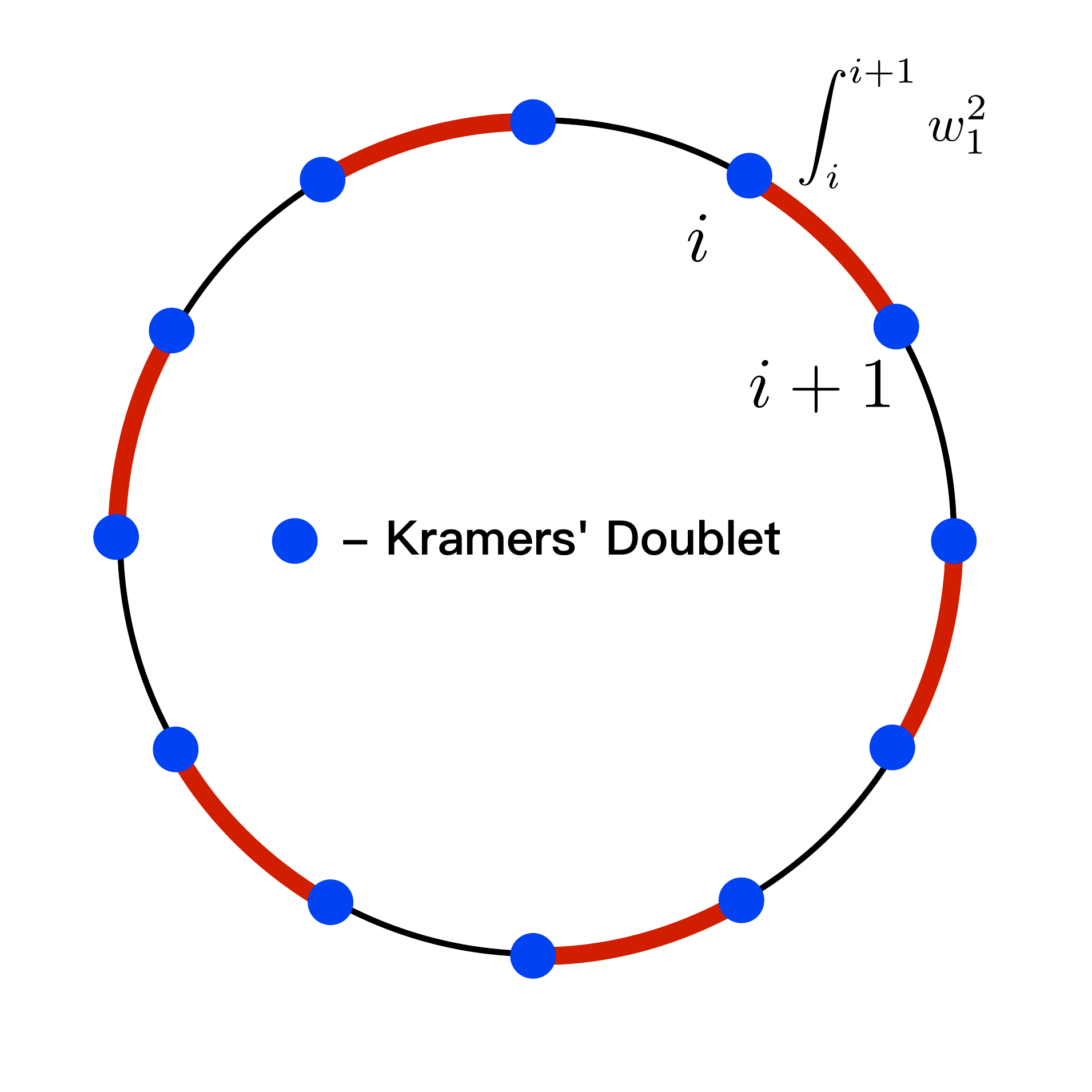} 
   \caption{A cartoon of a spin chain. }
   \label{fig:lattice}
\end{figure}

In this paper we do not dwell on the microscopic theories. However we would like to take a moment to comment on the appearance of the anomaly in the underlying lattice system. If we take the lattice to consist of spin-half particles on sites, then each site has a double-degeneracy (Kramers' degeneracy) as long as time-reversal is a good symmetry (denoted by blue dots in Fig.~\ref{fig:lattice}). Now we note that such a system can be though of as SPT phases (red lines) protected by time reversal symmetry, which end on the blue dots. Such an SPT phase in 1+1D was discussed in \cite{Kapustin:2014zva}, and its action must include the term $i \pi \int w_1^2$ on the over every other link of the chain\footnote{We have a path-integral picture of the chain in mind. In other words, the 1D chain is supplemented with the time direction. The integral $\int_i^{i+1} w_1^2$ between the link $i$ and $i+1$ is then interpreted as the integral over all time, and bounded by the spatial positions of $i$ and $i+1$ vertex.}. We now observe that translation symmetry (which maps to the $C$-symmetry in the continuum) generates exactly the change in the action
\be
\Delta S=i \pi \int w_1^2
\ee
which is precisely the same as \eqref{eq:anomaly_polynomial} discussed as in the context of the continuum. We therefore conclude that the system has the microscopic system with the same type of anomaly as its effective theory, as expected.

We have already discussed how the consistency of the XYZ model with the anomaly. However in this case the $SO(3)$ symmetry is not fully broken, but is reduced to the $K_4=\mathbb Z_2\times \mathbb Z_2$, which is still 't~Hooft anomalous with translation symmetry. To truly test the predictions of the \CxPxT anomaly, we would like to break the $SO(3)$ symmetry completely. We suggest the spin chain with the Hamiltonian
\be
H=H_{\mathrm{XYZ}}+\sum_{i=1}^{L}\sum_{a,b=x,y,z}K_{ab}\tau_{a,i}\tau_{b,i+2}, 
\ee
where  $K_{ab}=K_{ba}$, which we call the skewed-XYZ model\footnote{Notice that both first and second term leave invariant a 't Hooft anomalous $K_4\subset SO(3)$, and only the combination of the two break the $SO(3)$ symmetry completely.}. It becomes the XYZ model when $K_{ab}=0$, but it completely breaks $SO(3)$ in generic cases.

\section{Conclusion}
\label{sec:conclusion}

We have discussed here the appearance of the mixed 't~Hooft anomaly between charge conjugation $C$, parity $P$, and time-reversal $T$ symmetries for the two-dimensional $O(3)$ nonlinear sigma model at $\theta=\pi$, as well as a general Abelian-Higgs model with two scalar fields. 
Since it does not involve the $SO(3)$ symmetry, the anomaly matching constraint for this system exists even if we introduce terms to break the $SO(3)$ symmetry completely, so long as $C$, $P$, and $T$ are good symmetries of the perturbation. 
As a consequence of anomaly matching, the system cannot have the trivial gapped phase, and thus it must break $C$, $P$, or $T$ spontaneously, have massless excitations, or have nontrivial topological order.

The \CxPxT anomaly discussed here should be viewed as the generalization of the LSM theorem to spin-systems without the global $SO(3)$ spin symmetry. 
This \CxPxT anomaly is the field-theoretic description of the LSM theorem involving $T$ and the lattice symmetries, discussed in Ref.~\cite{chen2011,PhysRevB.93.104425}. 

An interesting future perspective would be to explore such anomalies in higher dimensional spin-half systems. In fact a hint that such an anomaly indeed exists is contained in the observation that a rectangular spin-1/2 system can be viewed as a series of spin-1/2 chains, each of which can only be realized as a boundary state of an SPT phase protected by the ``$C, P, T$'' symmetries discussed here. Therefore such a $2+1$D system  can be viewed a staggering of the 1+1D trivial and non-trivial SPT phases living between the chains. Such a setup breaks the translation symmetry shifting the stack of spin-1/2 chains, and implies that there should be an anomaly involving spatial reflections, translations and time reversal in $2+1D$ systems as well. If this is the case, then a similar argument can be made for $3+1$D spin systems on a cuboid lattice. We leave this interesting problem for the future. 

Another interesting extension of this work is to include the fermionic degrees of freedom, in particular extending the model to a supersymmetric one. The supersymmetric $O(3)$ nonlinear sigma model has a classical $U(1)$ chiral symmetry reduced to $\mathbb Z_2$ by instanton effects, and is believed to be spontaneously broken. Should gauging space-time symmetries (which should be possible unless discrete space-time symmetries are themselves 't Hooft anomalous) lead to fractional quantization of topological charge, the $\mathbb Z_2$ chiral symmetry would be eliminated completely, implying a mixed 't Hooft anomaly between discrete space-time symmetries and discrete the chiral symmetry.

\begin{acknowledgements}
We thank Gerald Dunne and Zohar Komargdoski for their useful comments on the initial draft of the paper. We also benefited greatly from the discussions with Costas Bachas, Gerald Dunne, Zohar Komargdoski, Anders W. Sandvik, and Mithat \"Unsal. 
We also appreciate Ken Shiozaki and Yohei Fuji drawing our attention to Refs.~\cite{chen2011, PhysRevB.93.104425}. 
The authors would also like to thank the hospitality at the KITP institute and the organizers of the ``Resurgent Asymptotics in Physics and Mathematics'' workshop where a part of this was done. The work at KITP is supported by National Science Foundation under Grant No. NSF PHY-1125915. 
The work of Y.~T. is supported by RIKEN Special Postdoctoral Researchers Program. 
\end{acknowledgements}

\begin{appendix}

\section{'t Hooft anomaly matching}\label{app:anomaly_matching}

In this appendix, we give a brief review on the anomaly matching condition. 

Let us consider quantum field theory on $d$-dimensional spacetime (i.e. $d=D+1$ with $D$ space dimension) with a global symmetry $G$. We introduce the background $G$-gauge field $A$ to the system, and denote the partition function as $Z[A]$, then the 't Hooft anomaly is defined as the gauge non-invariance on the phase of the partition function:
\be
Z[A+\delta_{\theta}A]=Z[A]\exp(\im S[\theta,A]). 
\ee
Here, $\theta$ is the gauge parameter, $\delta_{\theta}A$ is the gauge variation of $A$ by $\theta$, and $S$ is the $d$-dimensional local functional of $\theta$ and $A$. 

The working assumption, which can be checked in each concrete example, is that the 't Hooft anomaly can be canceled by anomaly inflow from the $(d+1)$-dimensional SPT phase protected by the symmetry $G$. Such SPT phase is characterized by $(d+1)$-dimensional action of the topological $G$-gauge theory $S_{\mathrm{SPT},d+1}[A]$,  and the assumption claims that 
\be
S[\theta,A]=\delta_{\theta}S_{\mathrm{SPT},d+1}[A], 
\ee
when the SPT phase has the $d$-dimensional boundary. As a result, the combined system, 
\be
Z[A]\exp(-\im S_{\mathrm{SPT},d+1}[A]), 
\ee
becomes $G$-gauge invariant. The SPT action cannot be influenced by the RG flow at the boundary, as it is a gapped phase and contains no propagating degrees of freedom.
In other words, whatever the low-energy effective theory of $Z[A]$ is, it must cancel the gauge non-invariance of $S_{\mathrm{SPT},d+1}$ coming out of the boundary, so the low-energy effective theory should have the same 't Hooft anomaly of the original theory. This is the 't Hooft anomaly matching condition. 

By taking the ultimately low-energy limit, we can claim that the anomaly should be matched by massless excitations or the degeneracy of the ground states since we can integrate out the massive excitations. 
The vacuum degeneracy may be created by the spontaneous symmetry breaking of a global symmetry group $G$ or the topological order. So a 't Hooft anomaly matching requires 
\begin{itemize}
\item spontaneous symmetry breaking of some part of $G$,
\item conformal behavior, or
\item topological order. 
\end{itemize}
The 't Hooft anomaly provides a nontrivial consequence on the nonperturbative dynamics of interacting quantum field theories. 

\section{$K_4-C$ mixed anomaly}\label{app:K4-C}

We here briefly discuss the mixed anomaly between $K_4=\mathbb Z_2\times \mathbb Z_2$ and the $C$ symmetry of the effective field theories of spin chains given by \eqref{eq:AH}. The group $K_4$ is the subgroup of the spin $SO(3)$ which is generated by $\pi$ rotations around the $x$ and $y$ axis of the spin. The consequences of this anomaly appeared in \cite{Jian:2017skd}. We here want to put the observations in the context of the 't Hooft anomaly matching. 

The group $K_4$ has four elements, which are the unit element along with $\pi$ rotations around the three axes N\'eel vector axes $x,y,z$, which we label as $Q_x,Q_y, Q_z$. The group elements act projectively on the $u$-field as elements $\bm 1, i\tau^1,i\tau^2,i\tau^3$.

Now let us think about adding the gauge field for the two $\mathbb Z_2$-s of the $K_4$. The associated $\mathbb Z_2$ gauge fields we take to be $A^x$ and $A^y$.  Notice that under the $C$-symmetry $A^x$ changes sign, while $A^y$ is invariant. 

\begin{figure}[tbp] 
   \centering
   \includegraphics[width=3in]{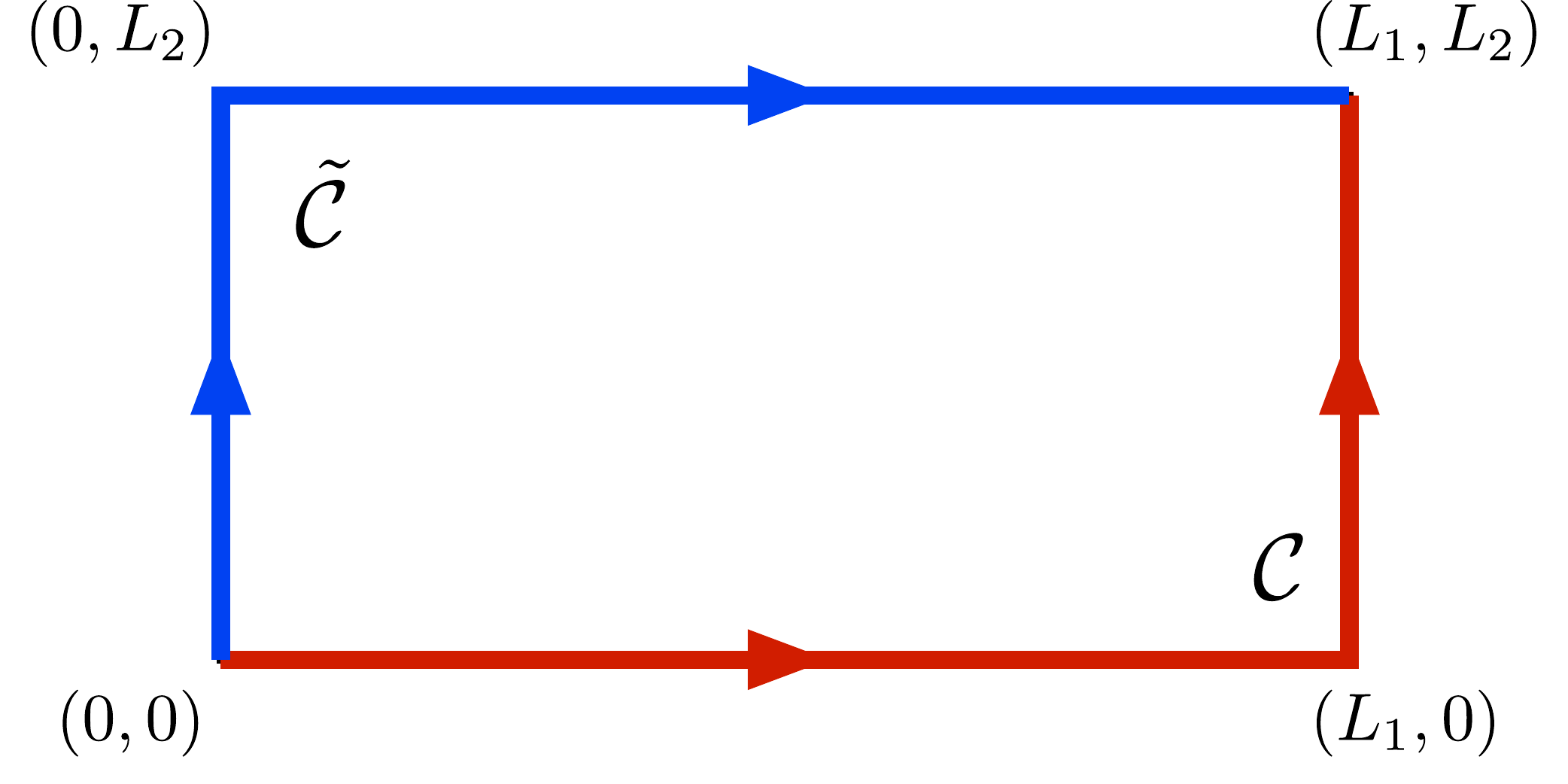} 
   \caption{The depiction of an open $2$-torus and contours. }
   \label{fig:example}
\end{figure}

Let us now think of adding the gauge field $A^x$ along the $1$-direction\footnote{We reserve the numbers $1,2$ for space-time indices, and $x,y,z$ for N\'eel vector $\bm$ directions.} and $A^y$ along the $2$-direction, so that $e^{i\int_{\gamma_1} A^x}=e^{i\int_{\gamma_2}A^y}=-1$, where $\gamma_{1,2}$ cycles. Think of the spatial $2$-torus as an open rectangle and allowing the $u$ field to obey the periodicity conditions
\begin{align}
u(L_1,x_2)=i\tau^1e^{i\int_{\mathcal C}A}u(0,x_2)\\
u(x_1,L_2)=i\tau^2e^{i\int_{\tilde{\mathcal C}}A}u(x_1,0)\;.
\end{align}
where we also inserted the open $U(1)$ Wilson loop so that the above conditions are properly gauge covariant. The contours $\mathcal C,\tilde{\mathcal  C}$ are such that the connect the points which are to be identified once the rectangle is wrapped back on the torus. Let us pick two contours depicted in Fig.~\ref{fig:example}.

We have that
\be
u(L_1,L_2)=-\tau_2 \tau_1 e^{i\int_{\mathcal C}A}u(0,0)=-\tau_1 \tau_2 e^{i\int_{\tilde{\mathcal C}}A}u(0,0)
\ee
which can only be true if
\be
e^{i \mathcal \int_{C\cup-\tilde{\mathcal C}}A}=e^{i\int_{\mathbb T^2} F[A]}=-1\;.
\ee
Therefore we find that the flux of the $U(1)$ gauge field on the $2$-torus must be $\pi$ modulo $2\pi$. 




We can also add a counter-term
\be\label{eq:S_counter_2D}
S_{\mathrm{counter}}=i\frac{p}{\pi}\int A^x\wedge A^y\;.
\ee
The above term is perfectly gauge invariant (modulo $2\pi i$) under $A^{x,y}\rightarrow A^{x,y}+d\varphi^{x,y}$ as long as $p\in \mathbb Z$, because $e^{2i\int_\gamma A^{x,y}}=1$ over any 1-cycle $\gamma$ since $A^{x,y}$ are $\mathbb Z_2$ gauge fields. If $A^x_1$ and $A^y_2$ are non-trivial, the above term contributes $ip\pi$ to the action, so any odd $p$ is equivalent to $p=1$ and any even $p$ is equivalent to $p=0$, so there is a unique gauge-invariant counter-term we can write. 

This term is invariant under the $C$-symmetry, and therefore cannot fix the problem with \eqref{eq:S_counter_2D}. Hence there is no local counter-term allowed which can fix the $C$-symmetry.

An attempt to gauge the $C$-symmetry would force us to attach a surface integral of the type \eqref{eq:S_counter_2D} to a stitch which implements a $C$-transformation. The only way to do this is to extend the background gauge fields into the bulk of a 2-torus and define the $C$ transformation to be attached to a 2-surface in the bulk. 

We can summarize this feature with an anomaly polynomial
\be\label{eq:2D_anomaly_polynomial}
\frac{i}{\pi^2}\int_{\Sigma}A_C \wedge A^x\wedge A^y\;,
\ee
where $A_C$ is the $\mathbb Z_2$ gauge field for the $C$-symmetry, and the integral is over the space $\Sigma$ whose boundary is the 2-torus, i.e. $\partial \Sigma=\mathbb T^2$. 
That is, this topological action characterizes the bulk SPT phase used for the 't Hooft anomaly matching argument in Appendix~\ref{app:anomaly_matching} for $K_4-C$ anomaly. 

\section{Symmetries: from Hilbert space to path-integrals}\label{app:Noether}

\subsection{Unitary symmetries}

To start we first define a Euclidean time matrix element
\be\label{eq:eu_mat_el}
\braket{\psi}{e^{-\beta H}}{\chi}\;.
\ee
To represent the matrix element as the path integral, we typically choose a particular complete basis of states, typically coordinate eigenstates, which we will denote as $\ket{\phi}$. Note that the state itself may depend on space --- i.e. an index labeling coordinates -- but since we are working in the Heisenberg picture (in Euclidean time) the state has no time dependence. We also have the completeness relation
\be
\int d\phi\; \ket{\phi}\bra{\phi}=\bm 1\;.
\ee
where $d\phi$ is the appropriate measure on the coordinate space. The matrix element \eqref{eq:eu_mat_el} can then be written as
\be 
\left(\prod_n\int d\phi_n\right) \brkt{\psi}{\phi_1}e^{-\sum_{n=1}^N\epsilon S(\phi_n,\phi_{n+1})}\brkt{\phi_N}{\chi}\;,
\ee
where $\epsilon=\beta/N$ and where\footnote{The ``link action" $S(\phi_n,\phi_{n+1})$ strictly depends also on $\epsilon$. However in the limit $\epsilon\to 0$, the dependence on $\epsilon$ dependence will be irrelevant, so we omit it for simplicity.}
\be
e^{-\epsilon S(\phi_n,\phi_{n+1})}\equiv \braket{\phi_n}{e^{-\epsilon H}}{\phi_{n+1}}\;.
\ee
Now let us do a symmetry transformation $\ket\psi,\ket\chi\rightarrow U\ket\psi,U\ket \chi$, where $U$ is a unitary operator. The matrix element \eqref{eq:eu_mat_el} is invariant, because $U$ commutes with the Hamiltonian. How is this invariance manifested in the path-integral. 

Let 
\be\label{eq:U_on_phi}
U^\dagger\ket\phi=\ket{\tilde\phi}
\ee
Then we have
\be
e^{-\epsilon S(\phi_n,\phi_{n+1})}= \braket{\phi_n}{Ue^{-\epsilon H}U^\dagger}{\phi_{n+1}}=e^{-\epsilon S(\tilde \phi_n,\tilde\phi_{n+1})}\;.
\label{eq:eq_mat_el}
\ee
Further, notice that the measure must be invariant $d\phi=d\tilde\phi$, as the completeness relation is preserved under the unitary transformation. 

Therefore we have that the total action
\be
S_{tot}=\sum_{n}\epsilon S(\phi_n,\phi_{n+1})
\ee
is invariant under the transformation which takes $\phi\rightarrow \tilde \phi$.

The arguments can be now reversed: if we assume that there exists a transformation of the coordinates such that the discretized action density (i.e. Lagrangian) $S(\phi_n,\phi_{n+1})=S(\tilde\phi_n,\tilde\phi_{n+1})$ is invariant, then we have that the transformation, defined by \eqref{eq:U_on_phi}, is a symmetry of the theory, i.e. that this symmetry operator commutes with the Hamiltonian. For continuum symmetries, this is the standard Noether theorem.

\subsection{Anti-unitary symmetries}

Time reversal symmetry is an anti-unitary (and hence anti-linear) operator. A crucial property of the anti-linear operator is that it complex-conjugates the coefficients of the basis states, while it acts as a linear operator on the basis states themselves. This however means that we must specify the direction of its action. We do this by drawing a small arrow on top. If the arrow is omitted it means that the operator acts to the right. Furthermore we have that if $A$ is the anti-linear operator, then
\be
\braket{\psi}{\overrightarrow A}{\chi}=\braket{\psi}{\overleftarrow A}{\chi}^*
\ee
Furthermore an anti-unitary operator is such that it leaves the norm of any given state invariant
\be
\brkt{U\psi}{U\psi}=\braket{\psi}{{\overleftarrow U}^\dagger \overrightarrow U}{\psi}=\braket{\psi}{{U}^\dagger U}{\psi}^*=\brkt{\psi}{\psi}\;.
\ee
This means that
\be
U^\dagger U=\bm 1\;.
\ee

If we start with the matrix element as in \eqref{eq:eq_mat_el}, then we get that under an anti-unitary transformation the matrix element changes as
\be
\braket{\psi}{e^{-\beta H}}{\chi}\rightarrow \braket{U\psi}{e^{-\beta H}}{U\chi}=\braket{\chi}{e^{-\beta H}}{\psi}\;.
\ee
Note that the sign in front of the Euclidean time $\beta$ does not flip sign, however the in and out state do exchange. In contrast the real-time element
\be
\braket{\psi}{e^{-i TH}}{\chi}\rightarrow \braket{U\psi}{e^{-iT H}}{U\chi}=\braket{\chi}{e^{i T H}}{\psi}\;.
\ee

The reversal of the sign in front of the real-time and the exchange of the in and out state indicate is a feature of time-reversal. Indeed time reversal is an anti-unitary unitary operator precisely for this reason. Because a general anti-unitary operator flips the sign of the real time, in the remaining discussion we will call it as the time-reversal operator.

The matrix element itself is not time reversal invariant, so it transforms nontrivially under time-reversal. Let us therefore specialize to a diagonal matrix element, i.e. so that $\chi=\psi$. We have that the Euclidean matrix element is invariant, while the real-time matrix element is invariant up to the change $T\rightarrow -T$. 

Now we repeat the same discussion as in Section \ref{app:Noether}. We get that we can represent the Euclidean matrix element by
\bea
&&\braket{\psi}{e^{-\beta H}}{\psi}\nonumber\\
&=&\left(\prod_{n} \int d\phi_n\right)\brkt{\psi}{\phi_1}e^{-\sum_{n=1}^{N-1}S(\phi_n,\phi_{n+1})\epsilon}\brkt{\phi_N}{\psi}.\nonumber\\
\eea
We have that
\be
\brkt{\psi}{\phi_1}\rightarrow \brkt{U\psi}{\phi_1}=\brkt{\psi}{U^\dagger \phi_1}^*=\brkt{U^\dagger \phi_1}{\psi}
\ee
and similarly
\be
\brkt{\phi_N}{\psi}\rightarrow \brkt{\psi}{U^\dagger\psi_N}\;.
\ee
Now under the anti-unitary $U$ we have that the basis states transform as a unitary operator
\be
U\ket\phi=\ket{\tilde \phi}\;.
\ee
so the eigenvalue $\phi$ transforms to $\tilde\phi$. We now have that 
\bea
e^{-\epsilon S(\phi_n,\phi_{n+1})}&=&\braket{\phi_n}{e^{-\epsilon H}}{\phi_{n+1}} \nonumber\\
&=& \braket{\phi_{n}}{\overrightarrow{U}\overrightarrow{U}^\dagger e^{-\epsilon H}}{\phi_{n+1}}\nonumber\\
&=&\braket{\tilde\phi_{n+1}}{e^{-\epsilon H}}{\tilde\phi_{n}}\nonumber\\
&=&e^{-\epsilon S(\tilde\phi_{n+1},\tilde \phi_{n})}.
\eea
The above transformation indicates that the fields $\phi_n$, transform as
\be
\phi_{n}\rightarrow \tilde\phi_{N-n}\;.
\ee
Since in the continuum limit the field will become $\phi(t)$, where $t$ is a continuous time label replacing $n$, we have the identification
\be
\phi_n=\phi(\epsilon n)\;.
\ee
This means that the continuous time takes values in $t\in[0,\beta)$. The transformation is therefore given by
\be
\phi(t)\rightarrow \tilde\phi(\beta-t)\;.
\ee
We could also define the continuous time $t$ to take values on a symmetric interval $t\in[-\beta/2,\beta/2]$. Then the transformation law would be
\be
\phi(t)\rightarrow \tilde \phi(-t)\;.
\ee

Notice that while the Euclidean time $t$ flips sign under the anti-unitary symmetry transformation, the limits of its integration are left unchanged. This is in contrast to the real time evolution, where invariance of the matrix element is only achieved up to the appropriate change of the limits of integration.

\end{appendix}

\bibliographystyle{utphys}
\bibliography{./QFT,./refs}


\end{document}